\documentclass[11pt,twoside]{article}
\usepackage{asp2006}
\usepackage{psfig}
\usepackage{epsf}
\usepackage{graphics}
\usepackage{lscape}

\usepackage{latexsym}
\usepackage{mathrsfs}
\usepackage{amsfonts}
\usepackage{amssymb}
\usepackage{amsmath}
\usepackage{natbib}
\usepackage{subfig}

\begin{document}
\title{The Parker Scenario for Coronal Heating as an 
MHD Turbulence Problem}

\author{A.F.~Rappazzo\altaffilmark{1},
M.~Velli\altaffilmark{1,2}, and
G.~Einaudi\altaffilmark{3}}

\altaffiltext{1}{Jet Propulsion Laboratory, California Institute of
                        Technology, Pasadena, CA, 91109 USA}

\altaffiltext{2}{Dipartimento di Astronomia e Scienza dello Spazio, 
                        Universit\`a di Firenze, 50125 Florence, Italy}

\altaffiltext{3}{Dipartimento di Fisica ``E.~Fermi'', Universit\`a di Pisa,
                        56127 Pisa, Italy}

\begin{abstract}
The Parker or field line tangling model of coronal heating is investigated through 
long-time high-resolution simulations of the dynamics of a coronal loop in cartesian
geometry within the framework of reduced magnetohydrodynamics (RMHD).  Slow photospheric motions induce a Poynting flux which saturates by driving an
anisotropic turbulent cascade dominated by magnetic energy and 
characterized by current sheets elongated along the axial magnetic field.
Increasing the value of the axial magnetic field different regimes
of MHD turbulence develop with a bearing on coronal heating rates.
In physical space magnetic field lines at the scale of convection cells
appear only slightly bended in agreement with observations of large loops
of current (E)UV and X-ray imagers.
\end{abstract}

\section{Introduction}

Coronal heating is one of the outstanding problems in solar physics.
Although the correlation of coronal activity with the intensity of 
photospheric magnetic fields seems beyond doubt, and there
is large agreement that photospheric motions are the source
of the energy flux that sustains an active region ($\sim 10^7\, erg\, cm^{-2}\, s^{-1}$),
the debate currently focuses on the physical mechanisms responsible for the
transport, storage and dissipation (i.e.\ conversion to heat and/or particle acceleration)
of this energy from the photosphere to the corona.

A promising model is that proposed by \citet{AR_park72,AR_park88},
who suggested that coronal heating could be the necessary outcome of
an energy flux associated with the tangling of coronal field
lines by photospheric motions.

Over the years numerous analytical and  numerical investigations
\citep{AR_park72, AR_StUch81, AR_vb86, AR_mik89, AR_park88, AR_ber91, 
AR_HP92, AR_GomFF92, AR_long94, AR_hen96, AR_ein96, AR_georg98, 
AR_dmi98, AR_dmi99, AR_ein99} have been carried out, 
discussing  in some details different aspects of the problem.

In all 3D cartesian simulations a complex coronal magnetic field results  
from the photospheric footpoints motions, and though the  field  does not, 
strictly speaking, evolve through a sequence of static force-free 
equilibrium states (the original Parker hypothesis), magnetic energy nonetheless
tends to dominate kinetic energy, and current sheets elongated along the axial 
direction characterize the system. The results from these studies agreed qualitatively among 
themselves, in that all simulations display the development of field aligned 
current sheets. However, estimates of the dissipated power and its scaling 
characteristics differed largely, depending on the way in which
extrapolations from low to large values of the plasma conductivity 
of the properties such as  inertial range power law 
indices were carried out. Furthermore, the physical mechanism responsible
for current sheets formation was not conclusively determined.

More recently a first attempt to simulate full 3D sections of the solar corona 
with a realistic geometry has been  performed by \citet{AR_gud05}. 
At the moment the very low resolution attainable with this kind of simulations
does not allow the development of turbulence. The transfer of energy 
from the scale of convection cells $\sim 1000\, km$
toward smaller scales is in fact inhibited, because the smaller scales are not resolved
(their linear resolution is in fact $\sim 500\, km$).

While in the future these global simulations will be able to reach the necessary
high resolutions, to investigate the nonlinear dynamics of the Parker
scenario at relatively high Reynolds numbers, we have recently
performed high-resolution long-time simulation of the aforementioned
cartesian model (\citet{AR_rap07}).

In the next sections we describe the coronal loop model, the simulations 
we have carried out, and give simple scaling arguments to understand 
the energy spectral slopes.

\section{Physical model}

A coronal loop is a closed magnetic structure threaded by a  strong axial 
field, with the footpoints rooted in the photosphere.
This makes it a strongly anisotropic system, as measured by 
the relative magnitude of the Alfv\'en velocity associated with
the axial magnetic field $v_{A} \sim 2000\ \textrm{km}\, \textrm{s}^{-1}$ 
compared to the typical photospheric velocity
$u_{ph} \sim 1\ \textrm{km}\, \textrm{s}^{-1}$. 
This means that the relative amplitude of the
Alfv\'en waves that are launched into the corona is very
small and, as an efficient energy cascade takes place \citep{AR_rap07},
the relative amplitude of the fields which develop in the
orthogonal planes remains small compared to the dominant
axial magnetic field.  

We study the loop dynamics  in a simplified cartesian geometry, 
neglecting any curvature effect,  as a ``straightened out'' box,  with an 
orthogonal square cross section of size $\ell$ (along which the x-y 
directions lie), and an axial length $L$ (along the z direction)
embedded in an axial homogeneous uniform magnetic field 
$\boldsymbol{B}_0 = B_0\ \boldsymbol{e}_z$. 
This simplified geometry allows us to perform simulations
with both high numerical resolution and long-time duration.

The dynamics of a plasma embedded in a strong axial magnetic field are
well described by the equations of reduced MHD \citep{AR_kp74, AR_stra76, 
AR_mont82}. 
In this limit the velocity and magnetic fields have only perpendicular components, 
linked to the velocity and magnetic potentials $\varphi$ and $\psi$ by
\begin{equation} \label{eq:potfi}
\boldsymbol{u}_\perp = \boldsymbol{\nabla} \times 
\left( \varphi\, \boldsymbol{e}_z \right), \qquad
\boldsymbol{b}_\perp = \boldsymbol{\nabla} \times 
\left( \psi\, \boldsymbol{e}_z \right).
\end{equation}
Although numerically we advance the equations for the potentials 
in order to analyze the linear and nonlinear properties of the system it is convenient 
to write the equivalent equations using the Els\"asser variables
$\boldsymbol{z}^{\pm} = \boldsymbol{u}_{\perp} \pm \boldsymbol{b}_{\perp}$.
The more symmetric equations, which explicit the underlying physical processes at 
work, are given in dimensionless form by:
\begin{eqnarray}
& &\frac{\partial \boldsymbol z^+}{\partial t}  =
- \left(  \boldsymbol z^- \cdot \boldsymbol{\nabla}_{\perp} \right) \boldsymbol{z}^+
+ \frac{v_A}{u_{ph}} \frac{\partial \boldsymbol z^+}{\partial z} 
+ \frac{1}{\mathcal{R}} \boldsymbol{\nabla}^{2}_{\perp} \boldsymbol z^+
 - \boldsymbol{\nabla}_{\perp} P \label{eq:els1}\\
& &\frac{\partial \boldsymbol z^-}{\partial t}  =
- \left(  \boldsymbol z^+ \cdot \boldsymbol{\nabla}_{\perp} \right) \boldsymbol{z}^-
- \frac{v_A}{u_{ph}} \frac{\partial \boldsymbol z^-}{\partial z}
+ \frac{1}{\mathcal{R}}   \boldsymbol{\nabla}^{2}_{\perp} \boldsymbol z^-
 - \boldsymbol{\nabla}_{\perp} P \label{eq:els2}\\
& &\boldsymbol{\nabla}_{\perp} \cdot \boldsymbol{z}^{\pm} = 0 \label{eq:els3}
\end{eqnarray}
where $P = p + \boldsymbol{b}_{\perp}^2/2 $ is the total pressure, and is
linked to the nonlinear terms by incompressibility~(\ref{eq:els3}):
\begin{equation}  \label{eq:els4}
\boldsymbol{\nabla}_{\perp}^2 P =
- \sum_{i,j=1}^2 \Big( \partial_i z_j^- \Big) \Big( \partial_j z_i^+ \Big).
\end{equation}

 The gradient operator has only components in the $x$-$y$ plane 
perpendicular to the axial direction $z$, and the dynamics in the 
orthogonal planes is coupled to the axial direction
through the linear terms $\propto \partial_z$.

We use a computational box with an aspect ratio of 10, which then spans
\begin{equation}
0 \le x, y \le 1, \qquad 0 \le z \le 10.
\end{equation}

The linear terms $\propto \partial_z$ are multiplied by the dimensionless
parameter $v_{A}/u_{ph}$, the ratio between 
the Alfv\'en velocity associated with the axial magnetic field
$v_{A} = B_0 / \sqrt{4 \pi \rho_0}$, 
and the photosperic velocity $u_{ph}$.

Boundary conditions for our numerical simulations are
specified imposing the velocity potential $\varphi (x,y)$ in the bottom
($z=0$) and top ($z=L$) planes:
\begin{equation}
\varphi (x,y) = \frac{1}{\sqrt{ \sum_{m,n} \alpha_{mn}^2 }}\,
\sum_{k,l} \frac{\alpha_{kl}}{2\pi \sqrt{k^2+l^2}}\,
\sin \left[ 2\pi \left( kx+ly \right) + 2\pi \xi_{kl} \right].
\end{equation} 
These result from the linear combination of large-scale eddies
with random amplitudes $\alpha_{kl}$ and phases $\xi_{kl}$ (whose
values are included between 0 and 1).
We excite all the twelve independent modes whose wave-numbers are 
included in the range $3 \le \left( k^2 + l^2 \right)^{1/2} \le 4$, 
and then normalize the result so that the velocity rms is 
$\sim 1\, km\, s^{-1}$.

In terms of the Els\"asser variables $\boldsymbol{z}^{\pm}$, to impose a
velocity pattern ($\boldsymbol{u}_{\perp}^*$) at the boundary surfaces means 
to impose the constraint
$\boldsymbol{z}^{+} + \boldsymbol{z}^{-} = 2 \boldsymbol{u}_{\perp}^*$,
and as in terms of characteristics (which in this case are simply $\boldsymbol{z}^{\pm}$
themselves) we can specify only the incoming wave
(while the outgoing wave is determined by the dynamics inside the computational
box), at the top ($z=L$) and bottom ($z=0$) planes the following 
``reflection'' takes place:
\begin{equation} \label{eq:bc0}
\boldsymbol{z^{-}}  = - \boldsymbol{z^{+}} + 2 \, \boldsymbol{u}^{0}_{\perp}
\quad \textrm{at} \ z=0
\end{equation}
\begin{equation} \label{eq:bcL}
\boldsymbol{z^{+}} = - \boldsymbol{z^{-}} + 2 \, \boldsymbol{u}^{L}_{\perp}
\quad \textrm{at} \ z=L
\end{equation}
where $\boldsymbol{u}^{0}_{\perp}$ and $\boldsymbol{u}^{L}_{\perp}$ are the forcing 
functions in the respective boundary surfaces.

At time $t=0$ no perturbation is imposed inside the computational box, 
i.e.\ $\boldsymbol{b}_{\perp} = \boldsymbol{u}_{\perp} = 0$, and
only the axial magnetic field $B_0$ is present: the subsequent dynamics are then
the effect of the photospheric forcing on the system. 

The linear terms ($\propto \partial_z$) in equations~(\ref{eq:els1})-(\ref{eq:els2})
give rise to two distinct wave equations for the $\boldsymbol{z}^{\pm}$ fields,
which describe Alfv\'en waves propagating along the axial direction $z$. 
This wave propagation, which is present during both the linear and nonlinear stages,
is responsible for the transport of energy at the large perpendicular scales
from the boundaries (photosphere) into the loop.
The nonlinear terms 
$\left(  \boldsymbol z^{\mp} \cdot \boldsymbol{\nabla}_{\perp} \right) \boldsymbol{z}^{\pm}$
are then responsible for the transport of this energy from the large scales toward the
small scales, where energy is finally dissipated, i.e.\ converted to heat and/or
particle acceleration.

An important feature of the nonlinear terms in 
equations~(\ref{eq:els1})-(\ref{eq:els3}) is the absence of self-coupling,
i.e.\ they only couple counterpropagating waves, and if one of the two fields
$\boldsymbol{z}^{\pm}$ were zero, there would be no nonlinear dynamics at all.
This is at the basis of the so-called Alfv\'en effect \citep{AR_iro64, AR_kra65}, 
that ultimately 
renders the nonlinear timescales longer and slows down the dynamics.

From this analysis it is clear that three different timescales are present: 
$\tau_{A}$, $\tau_{ph}$ and $\tau_{nl}$.
$\tau_{A} = L/v_{A}$ is the crossing time
of the Alfv\'en waves along the axial direction $z$, i.e.\
the time it takes for an Alfv\'en wave to cover the loop length $L$.
$\tau_{ph} \sim 5~m$ is the characteristic time associated with photospheric
motions, while $\tau_{nl}$ is the nonlinear timescale.

For a typical coronal loop $\tau_{A} \ll \tau_{ph}$, 
and for this reason we consider a forcing which is constant in time, i.e.\ 
for which formally $\tau_{ph} = \infty$.

In the RMHD ordering the nonlinear timescale $\tau_{nl}$ is bigger than the 
Alfv\'en crossing time $\tau_{A}$. This ordering is confirmed and
maintained troughout our numerical simulations.

The length of a coronal section is taken as the unitary length,
but as we excite all the wavenumbers between 3 and 4, and the
typical convection cell scale is $\sim 1000\, km$, this implies
that each side of our  section is roughly $4000\, km$ long.
Our grid for the cross-sections has 512x512 grid points,
corresponding to $\sim 128^2$ points per convective cell,
and hence a linear resolution of $\sim 8~km$.

Between the top and bottom plate a uniform magnetic field
$\boldsymbol{B} = B_0\, \boldsymbol{e}_z$ is present. The subsequent
evolution is due to the shuffling of the footpoints of the magnetic
field lines by the photospheric forcing.

In this section we present the results of a simulation performed with a numerical
grid with 512x512x200 points, Reynolds number  $\mathcal{R} = 800$, and the 
Alfv\'en velocity 
$v_{A}  = 200\, km\, s^{-1}$ corresponding to a ratio $v_{A}/u_{ph} = 200$.
The total duration is roughly 500 axial Alfv\'en crossing times
($\tau_{A} = L / v_{A}$).
\begin{figure}
      \plottwo{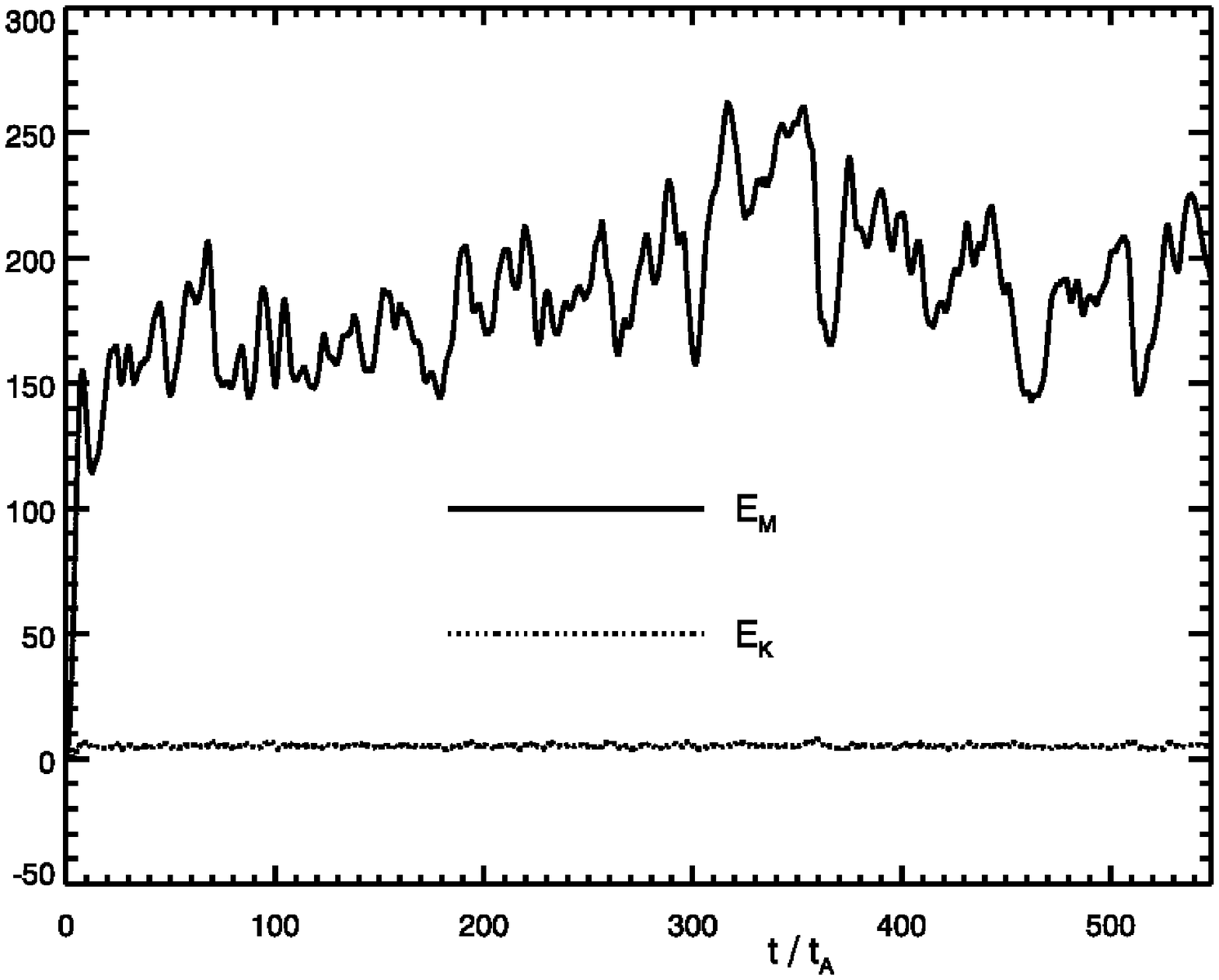}{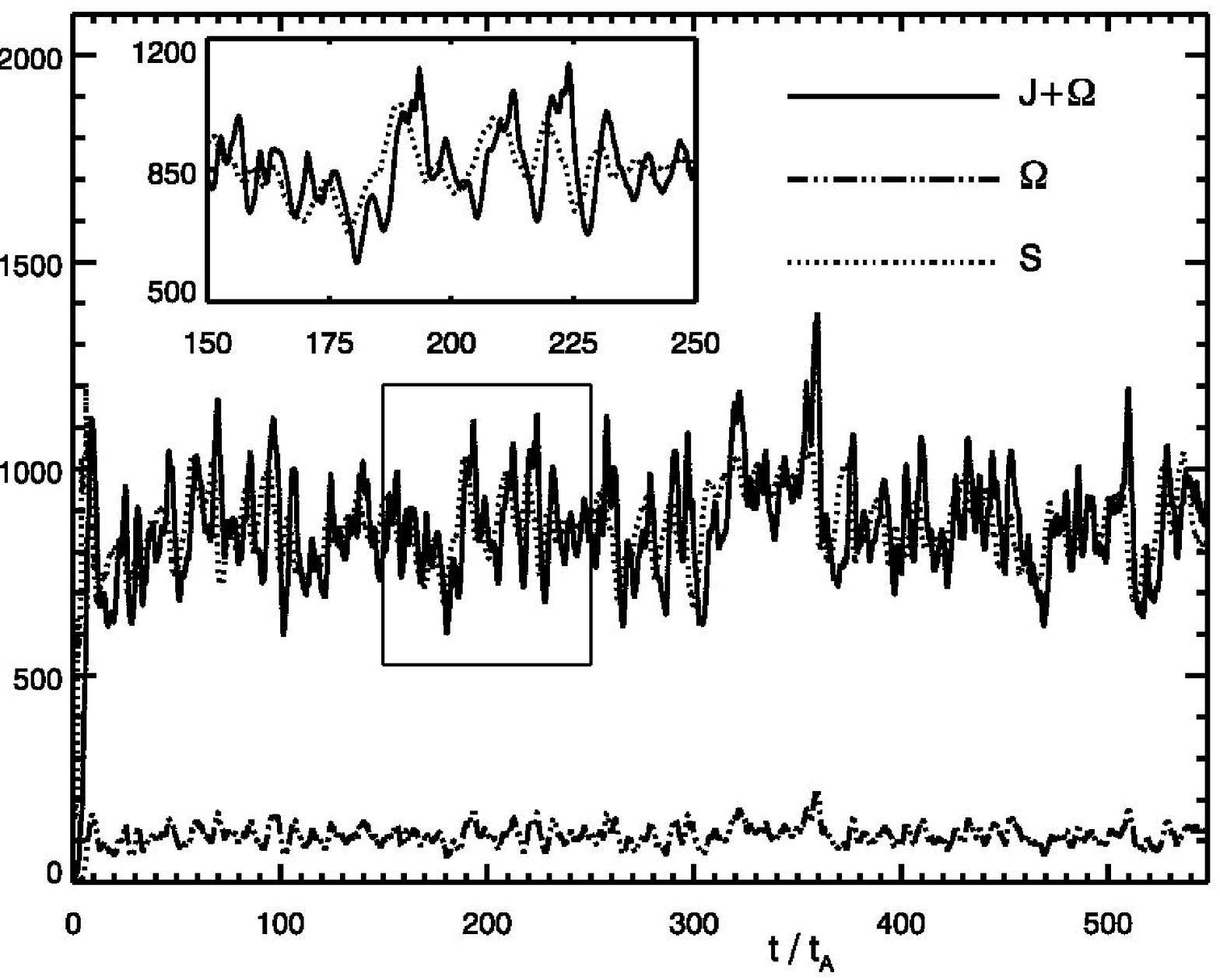}
      \caption{High-resolution simulation with
               $v_{A}/u_{ph} = 200$, 512x512x200
               grid points and $\mathcal{R} = 800$.
               \emph{Left}: Magnetic ($E_M$) and kinetic ($E_K$) energies as a function
               of time ($\tau_{A}=L/v_{A}$ is the axial Alfv\'enic
               crossing time). 
               \emph{Right}: The integrated Poynting flux $S$ dynamically balances the
               total dissipation $D$.
               Inset shows a magnification of total dissipation and
               $S$ for $180 \le t/\tau_{A} \le 280$.
      \label{fig:endiss}}
\end{figure}

Plots of the total magnetic and kinetic energies
\begin{equation}
E_M = \frac{1}{2} \int\! \mathrm{d} V\, \boldsymbol{b}_{\perp}^2,
\qquad
E_K = \frac{1}{2} \int\! \mathrm{d} V\, \boldsymbol{u}_{\perp}^2,
\end{equation}
and of the total magnetic ($J$) and kinetic ($\Omega$) dissipation rates
\begin{equation}
J = \frac{1}{\mathcal{R}} \int\! \mathrm{d} V\, \boldsymbol{j}^2
\qquad
\Omega =  \frac{1}{\mathcal{R}} \int\! \mathrm{d} V\, \boldsymbol{\omega}^2
\end{equation}
along with the incoming energy rate (Poynting flux) $S$,
are shown in Figure~\ref{fig:endiss}. At the beginning the
system has a linear behavior,
characterized by a time linear growth rate for the magnetic energy,
the Poynting flux and the electric current, until time
$t \sim 6\, \tau_{A}$, when nonlinearity sets in.
The magnetic energy is bigger than the kinetic energy, this is
the natural result of the field line bending due to the 
photospheric motions both in the linear and nonlinear stages.
More formally this is a consequence of the fact that, while on the perpendicular
magnetic field no boundary condition is imposed, the velocity field
must approach the imposed boundary values at the photosphere both
during the linear and nonlinear stages.

After this time, in the fully nonlinear stage, a \emph{statistically steady state}
is reached, in which the Poynting flux, i.e.\ the energy that is entering
the system for unitary time, balances on time average the total dissipation
rate ($J+\Omega$). As a result there is no average accumulation of energy
in the box, beyond what has been accumulated during the linear stage,
and a detailed examination of the dissipation time series (see inset in
Figure~\ref{fig:endiss}) shows that the Poynting flux and total dissipations
are decorrelated around dissipation peaks.
\begin{figure}
     \plottwo{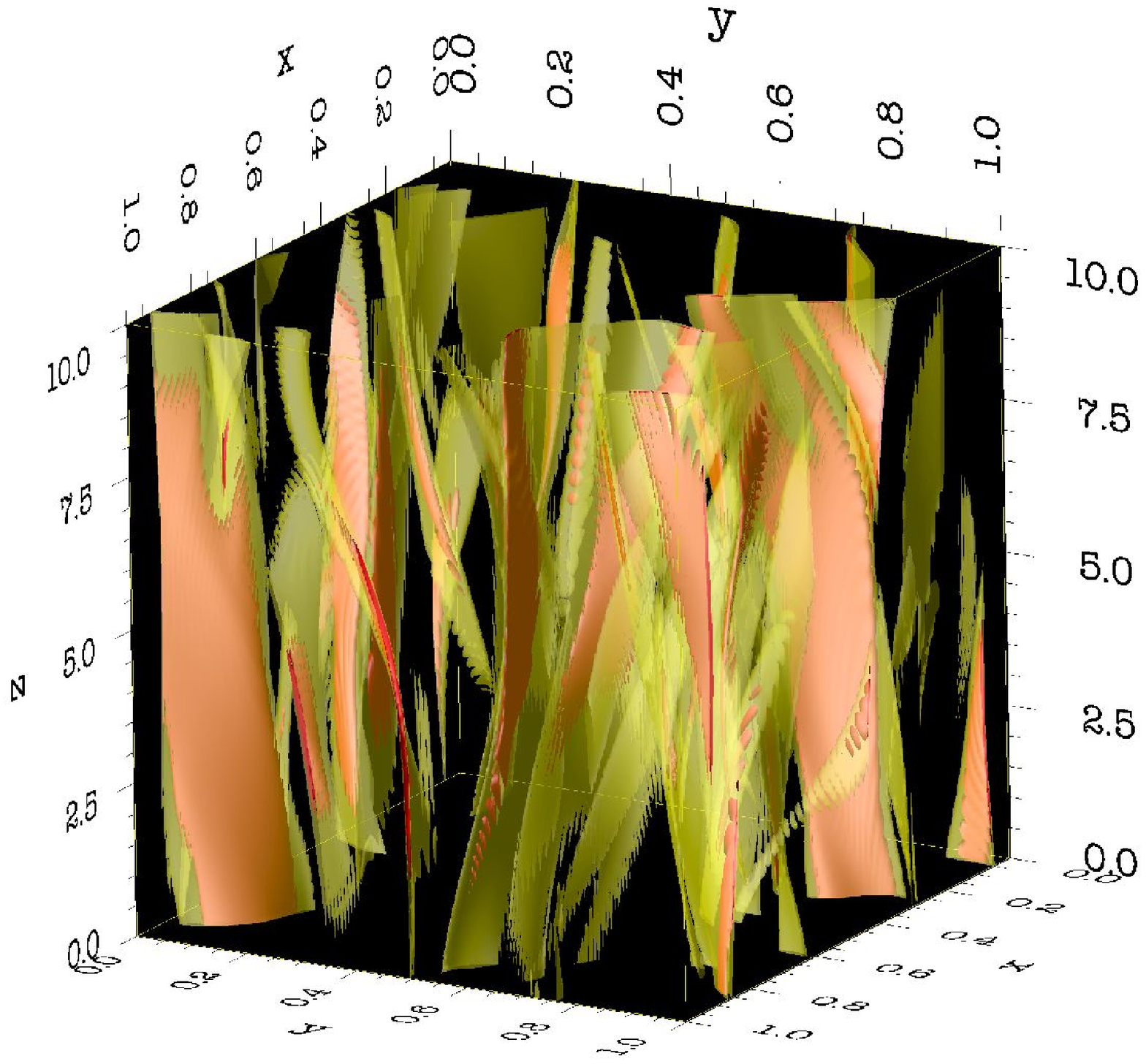}{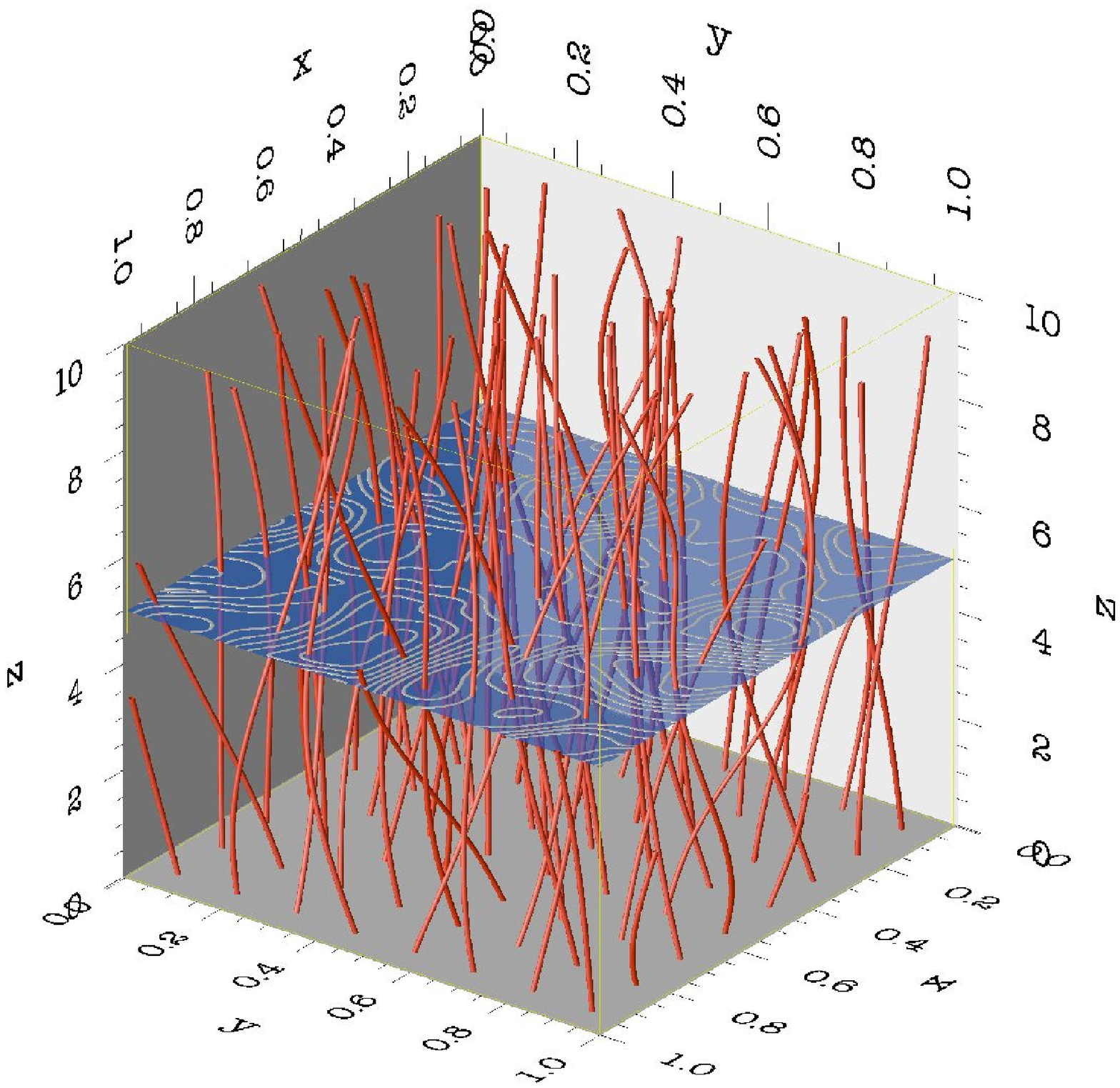}
      \caption{\emph{Left}: Snapshot of current sheets elongated along the axial direction.               \emph{Right}: Magnetic field lines are not much entangled because an efficient
      energy cascade takes place. Mid-plane shows field-lines of the orthogonal
      magnetic field.
      \emph{The box has been rescaled for an improved viewing. The box has
      an aspect ratio of 10, to restore the original aspect ratio the box should
      be stretched 10 times along the axial direction $z$.}
      \label{fig:ps}}
\end{figure}

Correspondingly, in Fourier space, we have shown \citep[see][]{AR_rap07} 
that the spectral index for total energy fits well the $-2$ value. We have 
also shown that this spectral index strongly 
depends on the ratio $v_A / u_{ph}$, i.e.\ on the relative
strength of the axial magnetic field. At lower values 
correspond flatter spectra, with an index close to $-5/3$ (Kolmogorov),
while to higher values of the magnetic field the spectra
steepens up to $\sim -5/2$ for $v_A / u_{ph} \sim 1000$.

Figure~\ref{fig:ps} shows a snapshot, at time $t \sim  18\, \tau_{\mathcal{A}}$,
of current sheets and field lines of total magnetic field.

\section{Conclusion and discussion}

Recently a lot of progress has been made in the understanding
of MHD turbulence for a system embedded in a strong
magnetic field, both in the condition of so-called weak 
\citep{AR_ng97, AR_gs97, AR_gal00} and strong turbulence
\citep{AR_gs95, AR_gs97, AR_cv00, AR_bisk00, AR_mbg03, AR_mg05}.
In particular it has been developed \citep{AR_bd05, AR_bd06, AR_mcb06} a
cascade model that self-consistently accounts for current sheets 
formation.

The presence of dynamical current sheets elongated along the direction
of the strong axial magnetic field, that are continuously formed
and dissipated, is most properly accounted for in the framework of MHD turbulence.
In fact, two important characteristic of MHD turbulence are that
the cascade takes place mainly in the plane orthogonal
to the local mean magnetic field \citep{AR_she83},
where small scales form, and that the small scales are not uniformly
distributed in this plane. Rather they are organized in 
current-vortex sheets aligned along the direction of the local 
main field, and \emph{constitute the dissipative structures of MHD
turbulence} (e.g.\ \cite{AR_bisk00}, \cite{AR_bisk03} and references therein). 
When the system is threaded by a strong axial 
magnetic field, as in our case, the cascade takes place mainly in the 
orthogonal planes, and the current sheets are elongated along
the axial direction (Figure~\ref{fig:ps}).

The fact that at the large orthogonal scales the Alfv\'en
crossing time $\tau_A$ is the fastest timescale, and in particular
it is smaller than the nonlinear timescale $\tau_{nl}$
(which can be identified with the energy transfer time
at the driving scale), implies that the Alfv\'en waves that continuously 
propagate and reflect from the boundaries toward the interior
are basically equivalent to an anisotropic magnetic forcing
function that stirs the fluid, whose orthogonal length is
that of the convective cells ($\sim 1000\, km$) and whose
axial length is given by the loop length $L$.

The spectra that we have found can be easily derived
by order of magnitude considerations.
Dimensionally, and integrating over the whole box, the 
energy cascade rate may be written as
\begin{equation} \label{eq:etr}
\epsilon \sim \ell_{\perp}^2\, L\, \rho\, 
\frac{{\delta z_{\lambda}}^2}{T_{\lambda}},
\end{equation}
where $\delta z_{\lambda}$ is the rms value of the Els\"asser fields
$\boldsymbol{z}^{\pm} = \boldsymbol{u}_{\perp} \pm \boldsymbol{b}_{\perp}$ at the 
perpendicular scale $\lambda$. Given the simmetry of the system 
it is expected and confirmed numerically that
cross helicity is zero, hence 
$\delta z^+_{\lambda} \sim \delta z^-_{\lambda} \sim \delta z_{\lambda}$. 
$\rho$ is the average density and 
$T_{\lambda}$ is the energy transfer time
at the scale $\lambda$, which is greater than the eddy turnover time 
$\tau_{\lambda} \sim \lambda / \delta z_{\lambda}$ because of the
Alfv\'en effect \citep{AR_iro64, AR_kra65}. 

In the classical IK case $T_{\lambda} \sim \tau_{\lambda} \, (\tau_{\lambda} / \tau_{A} )$.
More generally, however, as the Alfv\'en speed is increased nonlinear 
interactions become weaker. Simply from dimensional considerations
as the ratio $\tau_{\lambda}/\tau_{A}$ is dimensionless and smaller than 1,
we can suppose that the energy transfer time scales as 
\begin{equation} \label{eq:atnc}
T_{\lambda} \sim \tau_{\lambda} \, 
\left( \frac{\tau_{\lambda}}{\tau_{A}} \right)^{\alpha - 1}, 
\qquad \mathrm{with} \qquad \alpha \ge 1,
\end{equation}
where $\alpha$ is the scaling index
(note that $\alpha =1$ corresponds to standard hydrodynamic turbulence).

Substituting (\ref{eq:atnc}) in (\ref{eq:etr}) we can compute 
the energy transfer rate. As this is supposed to be constant along the
inertial range, considering the injection scale $\lambda \sim \ell_{\perp}$,
the energy transfer rate~(\ref{eq:etr}) becomes
\begin{equation} \label{eq:sbe2}
\epsilon 
\sim \ell_{\perp}^2 L\cdot \rho\, \frac{\delta z_{\ell_{\perp}}^2}{T_{\ell_{\perp}}} 
\sim \frac{\rho \ell_{\perp}^2 L^{\alpha}}{\ell_{\perp}^{\alpha} \, v_{A}^{\alpha - 1}} \, 
\delta z_{\ell_{\perp}}^{\alpha + 2}.
\end{equation}
On the other hand the energy injection rate is given by the 
Poynting flux integrated across the photospheric boundaries: 
$\epsilon_{in} = \rho\, v_{A} 
\int \! \mathrm{d} a\, \boldsymbol{u}_{ph} \cdot \boldsymbol{b}_{\perp}$.
Considering that  this integral is dominated by energy at the
large scales, due to the characteristics of the forcing function, we can approximate it with 
\begin{equation} \label{eq:pf}
\epsilon_{in} \sim \rho\, \ell_{\perp}^2 v_{A} u_{ph} \delta z_{\ell_{\perp}},
\end{equation}
where the large scale component of the magnetic
field can be replaced with $\delta z_{\ell_{\perp}}$ because the system is magnetically dominated.

The last two equations show that the system is self-organized because 
both $\epsilon$ and $\epsilon_{in}$ depend on $\delta z_{\ell_{\perp}}$, 
the rms values of the fields
$\boldsymbol{z}^{\pm}$ at the scale $\ell_{\perp}$:
the internal dynamics depends
on the injection of energy and the injection of energy itself depends 
on the internal dynamics via the boundary forcing. 

In a stationary cascade the injection rate (\ref{eq:pf}) is equal to 
the transport rate (\ref{eq:sbe2}). Equating the two determines
$\delta z_{\ell_{\perp}}$, that substituted  in (\ref{eq:sbe2}) or (\ref{eq:pf}) 
yields to the energy flux
\begin{equation} \label{eq:chs}
\epsilon^{\ast} 
\sim \ell_{\perp}^2 \, \rho \, v_{A} u_{ph}^2 
\left( \frac{\ell_{\perp} v_{A}}{L u_{ph}} \right)^{\frac{\alpha}{\alpha+1}},
\end{equation}
where $v_{A} = B_0 / \sqrt{4\pi\rho}$.
This is also the dissipation rate, and hence the \emph{coronal heating scaling}.
A dimensional analysis of eqs.~(\ref{eq:els1})-(\ref{eq:els3}) reveals
\citep[see][]{AR_rap07} that the only free parameter is 
$f = \ell_{\perp} v_{A}/ L u_{ph}$, so that 
the scaling index $\alpha$~(\ref{eq:atnc}), upon which the strength of 
the stationary turbulent regime depends, must be a function of $f$ itself,
and we have determined its value computationally \citep{AR_rap07}.

Dividing eq.~(\ref{eq:chs}) by the surface $\ell_{\perp}^2$ we obtain the energy flux 
per unit area $F=\epsilon^{\ast}/\ell_{\perp}^2$.
Taking for example a coronal loop $40,000\ \textrm{km}$ long, 
with a  number density of 
$10^{10}\ \textrm{cm}^{-3}$, $v_{\mathcal A} = 2,000\ \textrm{km}\, \textrm{s}^{-1}$ and  
$u_{ph} = 1\ \textrm{km}\, \textrm{s}^{-1}$, 
which models an active region loop, we obtain 
$F \sim 5 \cdot {10}^6\ \textrm{erg}\, \textrm{cm}^{-2}\, \textrm{s}^{-1}$. 
On the other hand, for
a coronal loop typical of a quiet Sun region, with a length of
 $100,000\ \textrm{km}$, a  number density of 
$10^{10}\ cm^{-3}$, $v_{\mathcal A} = 500\ \textrm{km}\, \textrm{s}^{-1}$ and  
$u_{ph} = 1\ \textrm{km}\, \textrm{s}^{-1}$, 
we obtain $F \sim 7\cdot {10}^4\ \textrm{erg}\, \textrm{cm}^{-2}\, \textrm{s}^{-1}$.

In summary,  the Parker scenario for coronal heating may
be considered as a MHD turbulence problem, self-consistently accounting for
current sheets formation. The derived coronal heating rates scale with coronal loop and 
photospheric driving parameters, and there is not a single universal scaling law.

\acknowledgements
A.F.R.\ is supported by the NASA Postdoctoral Program, M.V. is supported
by NASA LWS TR\&T and SR\&T.
A.F.R. and M.V. thank the IPAM program ``Grand Challenge Problems in
Computational Astrophysics'' at UCLA. \\


\begin{thebibliography}{}
\bibitem[Berger(1991)]{AR_ber91} Berger, M. A. 1991, A\&A, 252, 369

\bibitem[Biskamp(2003)]{AR_bisk03}Biskamp, D. 2003, Magnetohydrodynamic
Turbulence (Cambridge: Cambridge University Press)

\bibitem[Biskamp \& M\"uller(2000)]{AR_bisk00}Biskamp, D. \&
M\"uller, W.-C. 2000, Phys. Plasmas, 7, 4889 

\bibitem[Boldyrev(2005)]{AR_bd05}Boldyrev, S. 2005, \apj, 626, L37

\bibitem[Boldyrev(2006)]{AR_bd06}Boldyrev, S. 2006, 
Phys. Rev. Lett., 96, 115002

\bibitem[Cho \& Vishniac(2000)]{AR_cv00}Cho, J.
\& Vishniac, E.~T. 2000, \apj, 539, 273

\bibitem[Dmitruk et al.(1998)]{AR_dmi98}Dmitruk, P., 
G\'omez, D.~O. \& DeLuca, D.~D. 1998, \apj, 505, 974

\bibitem[Dmitruk \& G\'omez(1999)]{AR_dmi99}Dmitruk, P. \&
G\'omez, D.~O. 1999, \apj, 527, L63

\bibitem[Dmitruk et al.(2003)]{AR_dmi03}Dmitruk, P., 
G\'omez, D.~O. \& Matthaeus, W.~H. 2003, Phys. Plasmas, 10, 3584

\bibitem[Einaudi et al.(1996)]{AR_ein96}Einaudi, G., 
Velli, M., Politano, H. \& Pouquet, A. 1996, \apj, 457, L113

\bibitem[Einaudi \& Velli(1999)]{AR_ein99}Einaudi, G. \&
Velli, M. 1999, Phys. Plasmas, 6, 4146

\bibitem[Galtier et al.(2000)]{AR_gal00} Galtier, S., 
Nazarenko, S.~V., Newell, A.~C. \& Pouquet, A. 2000, J. Plasma Phys., 63, 447

\bibitem[Georgoulis et al.(1998)]{AR_georg98} Georgoulis, M.~K., 
Velli, M. \& Einaudi, G., 1998, \apj, 497, 957

\bibitem[Goldreich \& Sridhar(1995)]{AR_gs95}Goldreich, P.  \& Sridhar, S.
1995, \apj, 438, 763

\bibitem[Goldreich \& Sridhar(1997)]{AR_gs97}Goldreich, P.  \& Sridhar, S.
1997, \apj, 485, 680

\bibitem[Gomez \& Ferro-Fontan (1992)]{AR_GomFF92} Gomez,D.O. and Ferro-Fontan, C.F. 1992 \apj, 394, 662

\bibitem[Gudiksen \& Nordlund(2005)]{AR_gud05}Gudiksen, B.~V. \&
Nordlund, \AA. 2005, \apj, 618, 1020

\bibitem[Hendrix \& Van~Hoven(1996)]{AR_hen96}Hendrix, D.~L. \&
Van~Hoven, G. 1996, \apj, 467, 887

\bibitem[Heyvaerts \& Priest (1992)]{AR_HP92} Heyvaerts, J. and Priest, E.R. 1992, \apj, 390, 297 

\bibitem[Iroshnikov(1964)]{AR_iro64}Iroshnikov, P.~S. 
1964, Sov. Astron., 7, 566

\bibitem[Kadomtsev \& Pogutse(1974)]{AR_kp74}Kadomtsev, B.~B. 
\& Pogutse, O.~P. 1974, Sov. J. Plasma Phys., 1, 389

\bibitem[Kraichnan(1965)]{AR_kra65}Kraichnan, R.~H. 
1965, Phys. Fluids, 8, 1385

\bibitem[Longcope \& Sudan(1994)]{AR_long94}Longcope, D.~W. \&
Sudan, R.~N. 1994, \apj, 437, 491

\bibitem[Mason et al.(2006)]{AR_mcb06}Mason, J., Cattaneo, F.
\& Boldyrev, S. 2006, Phys. Rev. Lett., 97, 255002

\bibitem[M\"uller et al.(2003)]{AR_mbg03}M\"uller, W.~C., Biskamp, D.
\& Grappin, R. 2003, Phys. Rev. E, 67, 066302

\bibitem[M\"uller \& Grappin(2005)]{AR_mg05}M\"uller, W.~C. \& Grappin, R.
2005, Phys. Rev. Lett., 95, 114502

\bibitem[Mikic et al.(1989)]{AR_mik89}Mikic, Z., Schnack, D.~D. \&
Van~Hoven, G. 1989, \apj, 338, 1148

\bibitem[Montgomery(1982)]{AR_mont82}Montgomery, D.
1982, Phys. Scripta, T2/1, 83

\bibitem[Ng \& Bhattacharjee(1997)]{AR_ng97}Ng, C.~S. \&
Bhattacharjee, A. 1997, Phys. Plasmas, 4, 605

\bibitem[Parker(1972)]{AR_park72}Parker, E.~N. 1972, \apj, 174, 499

\bibitem[Parker(1988)]{AR_park88}Parker, E.~N. 1988, \apj, 330, 474

\bibitem[Rappazzo et al.(2007)]{AR_rap07}Rappazzo, A.~F., Velli, M.,
Einaudi, G. \& Dahlburg, R.~B. 2007, \apj, 657, L47

\bibitem[Shebalin et al.(1983)]{AR_she83} Shebalin, J.~V., Matthaeus, W.~H.
\& Montgomery, D. 1983, J. Plasma Phys., 29, 525

\bibitem[Sridhar \& Goldreich(1994)]{AR_sg94} Sridhar, S. \& Goldreich, P.
1994, \apj, 432, 612

\bibitem[Strauss(1976)]{AR_stra76}Strauss, H.~R. 1976, 
Phys. Fluids, 19, 134

\bibitem[Sturrock and Uchida (1981)]{AR_StUch81} Sturrock, P.A. and Uchida, Y., 1981, \apj, 246, 331

\bibitem[van Ballegooijen(1986)]{AR_vb86}van Ballegooijen, A.A. 1986, \apj, 311, 1001
\end{thebibliography}
\end{document}